\documentclass[a4paper,11pt]{article}
\usepackage{pos}

\usepackage{lipsum}
\usepackage{setspace}
\usepackage[capitalise]{cleveref}

\title{Astroparticle Physics with the Forward Physics Facility at the High-Luminosity LHC}
 \ShortTitle{Astroparticle Physics with the FPF at the HL-LHC}

\author*[a,b]{Dennis Soldin}

\affiliation[a]{\mbox{Karlsruhe Institute of Technology, Institute of Experimental Particle Physics, Karlsruhe, Germany}}
\affiliation[b]{University of Utah, Department of Physics and Astronomy, Salt Lake City, UT 84112, USA}

\emailAdd{soldin@kit.edu}

\abstract{
High-energy collisions at the High-Luminosity Large Hadron Collider (HL-LHC) will produce an enormous flux of particles along the beam collision axis that is not accessible by existing LHC experiments. Multi-particle production in the far-forward region is of particular interest for astroparticle physics. High-energy cosmic rays produce large particle cascades in the atmosphere, extensive air showers (EAS), which are driven by hadron-ion collisions under low momentum transfer in the non-perturbative regime of QCD. Thus, the understanding of high-energy hadronic interactions in the forward region is crucial for the interpretation of EAS data and for the estimation of backgrounds for searches of astrophysical neutrinos. The Forward Physics Facility (FPF) is a proposal to build a new underground cavern at the HL-LHC which will host a variety of far-forward experiments to detect particles outside the acceptance of the existing LHC experiments.

We will present the current status of plans for the FPF and highlight the synergies with astroparticle physics. In particular, we will discuss how measurements at the FPF will improve the modeling of high-energy hadronic interactions in the atmosphere and thereby reduce the associated uncertainties of measurements in the context of multi-messenger astrophysics. 
}

\ConferenceLogo{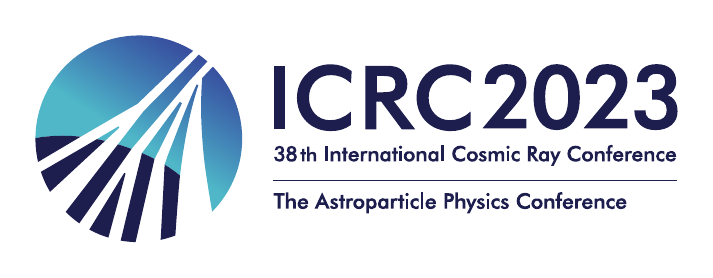}

\FullConference{%
38th International Cosmic Ray Conference (ICRC2023)\\
  26 July - 3 August, 2023\\
  Nagoya, Japan}

\begin{document}
\maketitle

\section{Introduction}

\begin{figure}[b]
    \vspace{-1.em}
    
    \centering
    \includegraphics[width=0.9\textwidth]{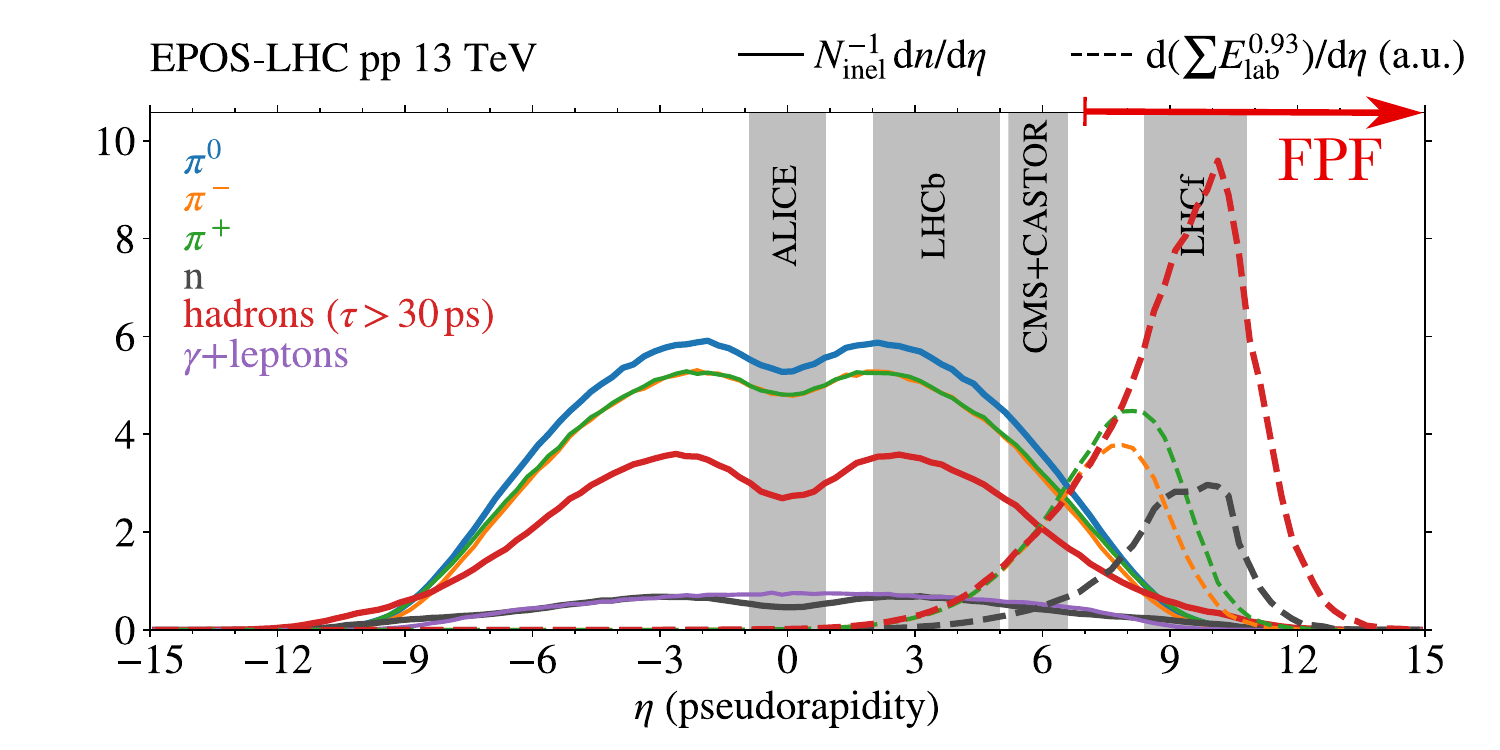}
    \vspace{-1.em}
    
    \caption{\label{fig:eta-ranges}Simulated densities of particles in arbritrary units (solid lines) in high-energy proton-proton collisions using EPOS-LHC. Dashed lines show the estimated number of muons produced by these particles if they were propagated through the atmosphere, assuming an equivalent energy for the fixed target collisions in the laboratory frame, $E_{\rm{lab}}$, and $N_\mu \propto E_{\rm{lab}}^{0.93}$. Figure taken from Ref.~\cite{Albrecht:2021cxw}.}
    \vspace{-0.5em}
\end{figure}

Throughout the history of particle physics, discoveries have been made through the observation of cosmic rays and neutrinos. Examples are the discovery of new elementary particles, the observation of neutrino oscillations, as well as measurements of cross-sections and particle interactions far beyond current collider energies. 
Cosmic rays enter the Earth’s atmosphere with energies exceeding $10^{11}~\rm{GeV}$, where they interact with molecules in the air. These collisions produce extensive air showers (EASs) in the atmosphere which can be measured with large detector arrays at the ground. \cref{fig:eta-ranges} shows simulated particle densities produced in proton-proton collisions (solid lines) compared to the pseudorapidity ($\eta$) ranges for current LHC experiments~\cite{Albrecht:2021cxw}. Also shown are the estimated number of muons, $N_\mu$, produced by these particles during propagation through the atmosphere, assuming $N_\mu \propto E_{\rm{lab}}^{0.93}$, where $E_{\rm{lab}}$ is the energy of the secondary EAS particles in the laboratory frame (dashed lines). While the mid-rapidity ranges are only marginally relevant for the particle production in EASs, the forward region ($\eta>4$) plays a crucial role for the EAS development.

Experimentally, the properties of the initial cosmic ray, such as energy and mass, are inferred indirectly from the particles measured at the ground by large detector arrays and their interpretation strongly relies on simulations of the EAS development. The main challenge in the description of EASs is the treatment of hadronic interactions in the forward region over many orders of magnitude in energy, which can not be probed by existing collider facilities. 
The EAS development is mainly driven by relativistic hadron-ion collisions in the atmosphere at low momentum transfer in the non-perturbative regime of quantum chromodynamics (QCD). Under these conditions, hadron production cannot be described using first principles and thus simulations rely on a variety of phenomenological hadronic interaction models. Large uncertainties remain due to those theoretical limitations and the lack of data from existing collider experiments. Hence, dedicated measurements of hadronic interactions under controlled experimental conditions, in particular in the far-forward region, are crucial in order to test and improve existing EAS models.

\section{The Forward Physics Facility}

The \emph{Forward Physics Facility} (FPF)~\cite{Anchordoqui:2021ghd,Feng:2022inv} is a proposal to build a new underground cavern at CERN to host a suite of far-forward experiments during the High-Luminosity LHC (HL-LHC) era. The existing LHC detectors have un-instrumented regions along the beam line and miss potential physics opportunities provided by the enormous flux of particles produced in the forward direction. In addition, without the FPF, the HL-LHC will be blind to neutrinos and many proposed new particles. However,  pathfinder experiments currently operating in the forward region at the LHC have recently directly observed collider neutrinos for the first time and demonstrated the potential for searches of new physics~\cite{FASER:2023zcr,SNDLHC:2023pun}. With the FPF, a diverse suite of experiments will realize this physics potential by detecting neutrino interactions at the highest human-made energies, expanding our understanding of particle interactions in the far-forward region.

\begin{figure}[bt]
\vspace{-1em}
\centering
\includegraphics[width=0.99\textwidth]{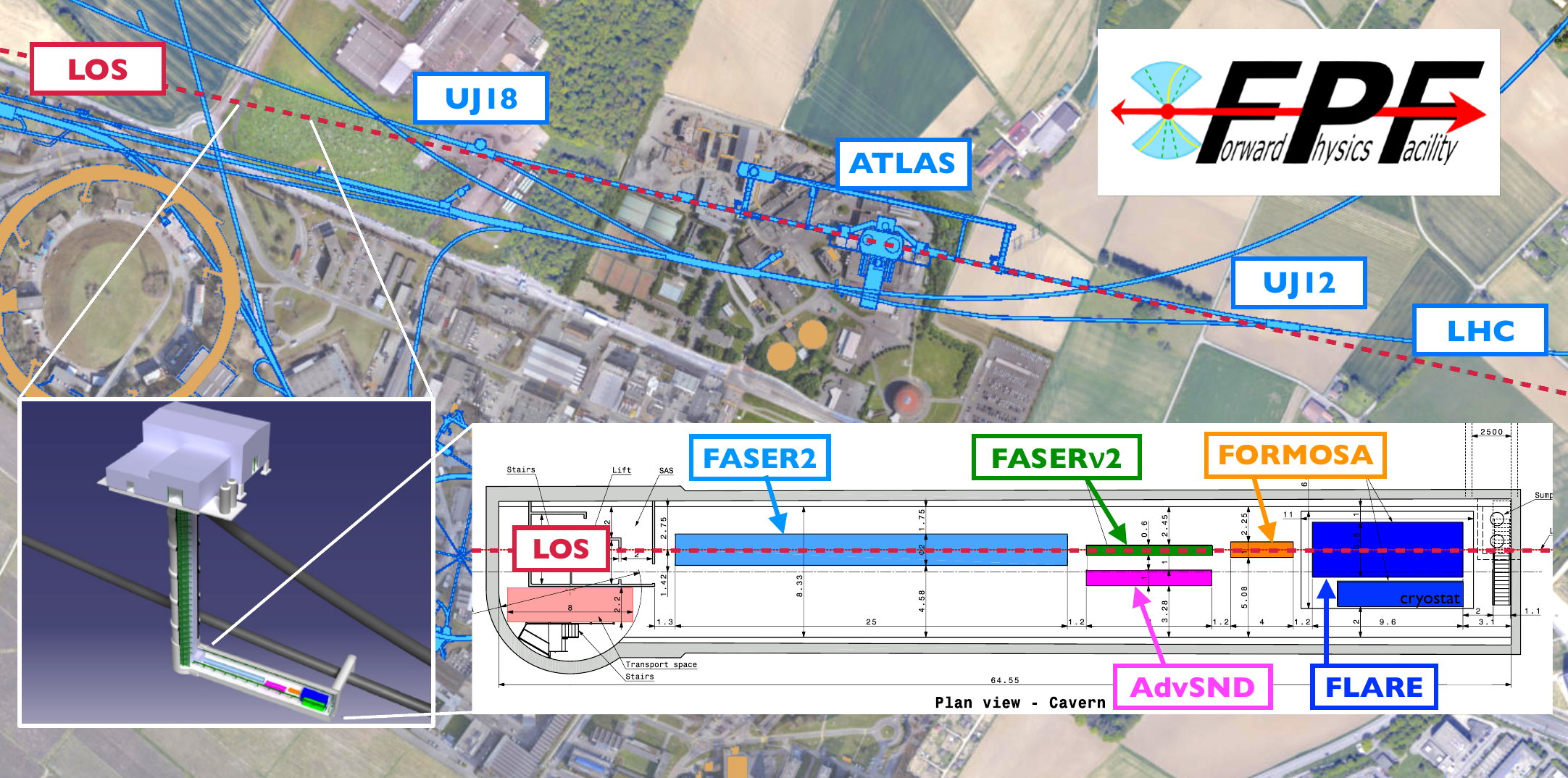}
\caption{\label{fig:ExecutiveSummaryMap}The preferred location for the Forward Physics Facility, a proposed new cavern for the High-Luminosity LHC era, on the collision
axis line-of-sight (LOS) of the ATLAS experiment. The FPF will be $65\,\rm{m}$-long and $8.5\,\rm{m}$-wide and will house a diverse set of experiments to explore the many physics opportunities in the far-forward region. Figure taken from Ref.~\cite{Feng:2022inv}.
}
\end{figure}

The proposed site at CERN is shown in \cref{fig:ExecutiveSummaryMap}. It is located at an depth of $88\,\rm{m}$, on the collision axis line-of-sight (LOS) of the ATLAS experiment, and $617\,\rm{m}$ to the west of the interaction point, shielded by more than $200\,\rm{m}$ of rock. The FPF will be $65\,\rm{m}$-long and $8.5\,\rm{m}$-wide and will house a diverse set of experiments to explore many physics opportunities at pseudorapidities above $\eta\sim 7$.

\subsection{Experiments}

Currently, five experiments are proposed to be housed at the FPF which are based on different detector technologies and optimized for particular physics goals~\cite{Feng:2022inv}. The proposed experiments are shown in \cref{fig:ExecutiveSummaryMap} and include:

\begin{itemize}
\setlength{\itemsep}{1pt}
\setlength{\parskip}{0pt}
  \setlength{\parsep}{0pt}
\item {\bf FASER2:} A magnetic tracking spectrometer, designed to search for light and weakly-interacting states, including new force carriers, sterile neutrinos, axion-like particles, dark sector particles, and to distinguish $\nu$ and $\bar\nu$ charged current scattering in the upstream detectors.
\item {\bf FASER$\nu$2:} An on-axis emulsion detector, with pseudorapidity range $\eta>8.4$, that will detect neutrinos at TeV energies with unparalleled spatial resolution, including tau neutrinos.
\item {\bf FLArE:} A noble liquid fine-grained time projection chamber to detect neutrinos and search for light dark matter with high kinematic resolution and wide dynamic range.
\item {\bf AdvancedSND:} An off-axis detector ($7.2 <\eta < 8.4$) to study neutrinos from charm decay and provide detailed observations of neutrino interactions for all neutrino flavors.
\item {\bf FORMOSA:} A detector composed of scintillating bars, with world-leading sensitivity to millicharged particles across a large range of masses.
\end{itemize}

This diverse suite of experiments will probe a large variety of physics, including new particles, neutrinos, dark matter, dark sectors, and QCD. For a detailed description of the individual experiments and their physics capabilities, please refer to Ref.~\cite{Feng:2022inv}. In this work, we will highlight the unique opportunities at the FPF for interdisciplinary studies at the intersection of high-energy particle physics and modern astroparticle physics.

\section{Synergies with Astroparticle Physics}

Measurements at the FPF will probe high-energy hadronic interactions in the far-forward region. These measurements will improve the modeling of high-energy hadronic interactions in the atmosphere, reduce the associated uncertainties of air shower measurements, and thereby help to understand the properties of the highest energetic cosmic rays. In addition, atmospheric neutrinos, produced in EASs in the far-forward region, are the main background for searches of high-energy astrophysical neutrinos with large-scale neutrino telescopes~\cite{IceCube:2016zyt,IceCube-Gen2:2020qha,Coniglione_2015}. Hence, measurements at the FPF will help to understand the atmospheric neutrino flux and reduce the associated uncertainties for astrophysical high-energy neutrino searches. These obvious connections between measurements at the FPF and modern astroparticle physics will be explored in the following.

\subsection{Light Hadron Production}
\label{sec:light_hadrons}
Muons are tracers of hadronic interactions and thus the measurement of muons in EASs is an important observable in order to test hadronic interaction models. Over the last 20 years, some EAS experiments reported discrepancies between model predictions and
experimental data, while other experiments reported no discrepancies~\cite{Albrecht:2021cxw}. The most unambiguous experimental evidence for a deficit of muons in simulations was observed in the analysis of data from the Pierre Auger Observatory~\cite{PierreAuger:2014ucz,PierreAuger:2016nfk}. A systematic meta-analysis of measurements of GeV muons from nine air shower experiments found an energy-dependent trend of the discrepancies with high statistical significance~\cite{Dembinski:2019uta,Soldin:2021wyv}, referred to as the \emph{Muon Puzzle} in EASs. Most recent studies indicate that the observed discrepancies may have a very complex nature which is currently not understood~\cite{ArteagaVelazquez:2023fda}.

The muons detected by air shower experiments are typically of low energy (a few to tens of GeV). They are produced at the end of a cascade of hadronic interactions with up to about 10 generations, where the dominant process is soft hadron production. As shown in \cref{fig:eta-ranges}, hadron production at forward pseudorapidities have the largest impact on muon production in EASs. The sensitivity to the produced hadrons is high and even small deviations of $\sim 5\%$ in the multiplicity and/or identity of the secondary hadrons have a sizeable impact on the muon flux~\cite{Baur:2019cpv}.

\begin{figure}[tb]
\centering
\vspace{-1em}
\includegraphics[width=.95\textwidth]{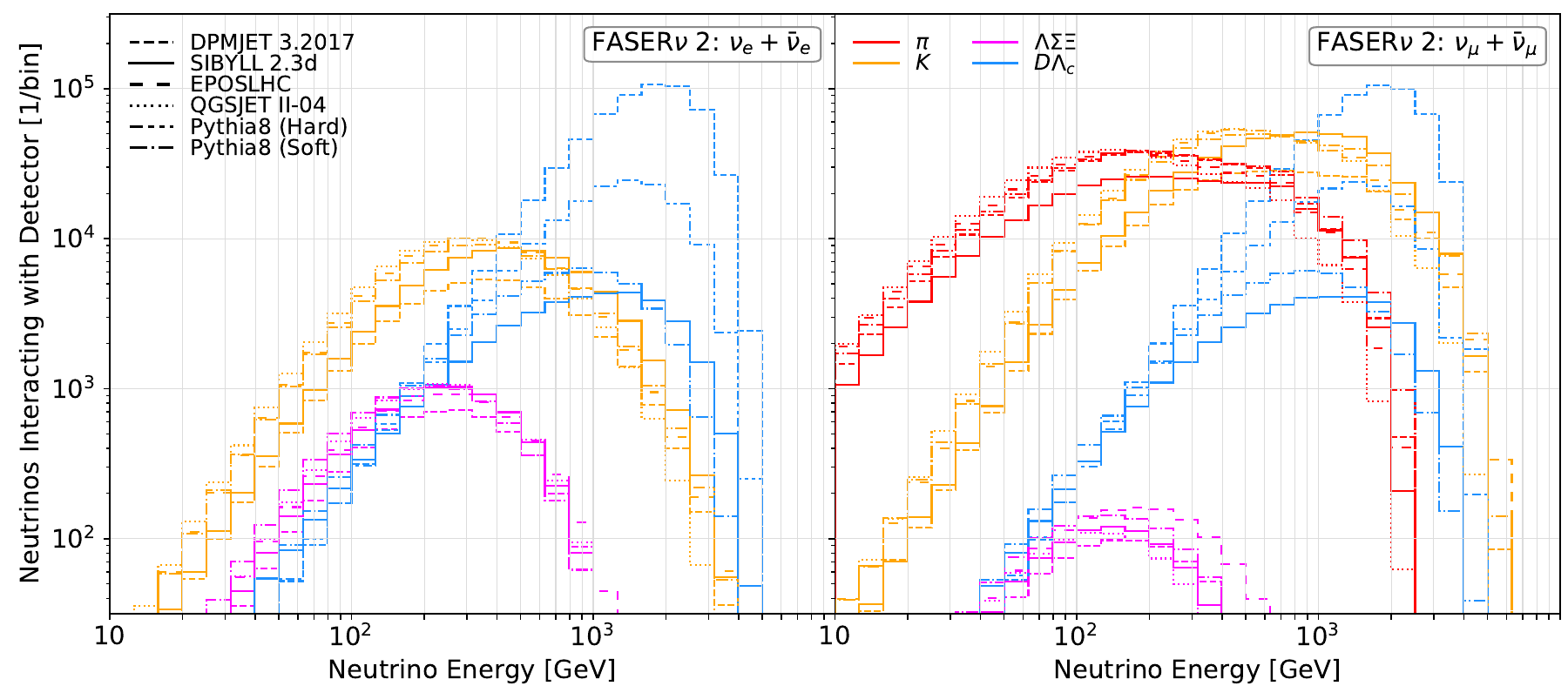}
\caption{\label{fig:nu-rate}Neutrino energy spectra for electron neutrinos (left) and muon neutrinos (right) passing through FASER$\nu$2 for an integrated luminosity of $3\,\rm{ab}^{-1}$. The different production modes are shown separately, i.e., pion decays (red), kaon decays (orange), hyperon decays (magenta), and charm decays (blue). The predictions are obtained from SIBYLL-2.3d (solid), DPMJET-III.2017 (short dashed), EPOS-LHC (long dashed), QGSJET-II.04 (dotted), and Pythia~8.2 using soft-QCD processes (dot-dashed) and with hard-QCD processes for charm production (double-dot-dashed). Figure taken from Ref.~\cite{Feng:2022inv}.}
\vspace{-1em}
\end{figure}

Dedicated measurements at the FPF will help to better understand light hadron production in the far-forward region. The ratio of electron neutrino ($\nu_e$) and muon neutrino ($\nu_\mu$) fluxes measured at the FPF experiments, for example, is a proxy for the ratio of charged kaons ($K$) to pions ($\pi$). Electron neutrino fluxes are a measurement of kaons, whereas both muon and electron neutrinos are produced via pion decay. However, $\nu_e$ and $\nu_\mu$ populate different energy regions, which can be used to disentangle them. \cref{fig:nu-rate} shows predictions of the neutrino energy spectra for $\nu_e$ and $\nu_\mu$ passing through FASER$\nu$2 for an integrated luminosity of $3\,\rm{ab}^{-1}$. The predictions are obtained from SIBYLL-2.3d~\cite{Riehn:2019jet}, DPMJET-III~\cite{Ranft:2002rj}, EPOS-LHC~\cite{Pierog:2013ria}, QGSJET-II.04~\cite{Ostapchenko:2019few}, and Pythia~8.2~\cite{Sjostrand:2014zea}. The resulting fluxes differ by up to a factor 2 for neutrinos from pion and kaon decays, which is much bigger than the anticipated statistical uncertainties at the FPF~\cite{Kling:2021gos}. 


\begin{figure}[tb]
\centering
\vspace{-1em}
\includegraphics[width=1.\textwidth]{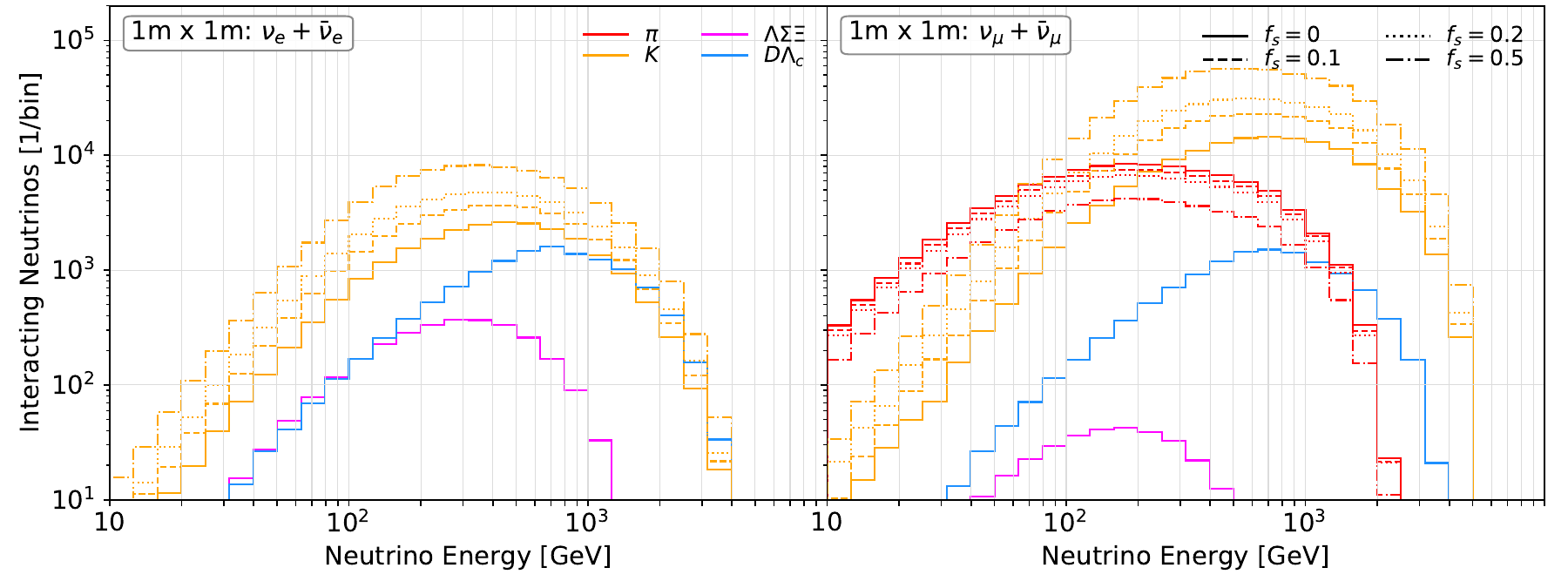}
\vspace{-1.2em}
\caption{\label{fig:dos}Neutrino energy spectra for electron neutrinos (left) and muon neutrinos (right) passing through the FLArE detector (10-ton target). The vertical axis shows the number of neutrinos that pass the detector's cross-sectional area of $1\,\mathrm{m}^2$ for an integrated luminosity of $3\,{\rm ab}^{-1}$: pion decays (red), kaon decays (orange), hyperon decays (magenta), and charm decays (blue). The different line styles correspond to predictions obtained from SIBYLL-2.3d by varying $f_s$~\cite{Anchordoqui:2022fpn}. Figure taken from Ref.~\cite{Feng:2022inv}.}
\end{figure}

A potential key to understand the Muon Puzzle is a universal strangeness and baryon enhancement observed by ALICE in collisions at mid-rapidity~\cite{ALICE:2016fzo}. This enhancement only depends on the multiplicity of the event and not on the details of the collision system, thus it allows to predict the hadron composition in EASs, in a phase space well beyond the reach of current colliders. If the enhancement increases in the forward region it will impact muon production in EAS and could be traced by the ratio of charged kaons to pions measured at the FPF. \cref{fig:dos} shows predictions obtained from a simple toy model based on SIBYLL-2.3d~\cite{Anchordoqui:2022fpn,Sciutto:2023zuz} in which the strangeness enhancement is realized by introducing the possibility of swapping $\pi\rightarrow K$ with a probability $f_s$ at large pseudorapidities, $|\eta|>4$. For $f_s = 0.1$ ($f_s = 0.2$) the predicted electron neutrino flux at the peak of the spectrum is a factor of 1.6 (2.2) larger than the baseline prediction. These differences are significantly larger than the anticipated uncertainties at the FPF. It has been shown that for $0.4 < f_s < 0.6$, which can be probed at the FPF, this simple model can partially accommodate data from the Pierre Auger Observatory~\cite{Anchordoqui:2022fpn} and thus provides a potential explanation for the Muon Puzzle. This example demonstrates how the FPF will provide unique tests of and constraints for hadronic interaction models to improve our understanding of multi-particle production in EASs.




\subsection{Charm Hadron Production}

High-energy neutrinos of astrophysical origin are routinely observed by neutrino telescopes and the next generation of neutrino telescopes, such as IceCube-Gen2~\cite{IceCube-Gen2:2020qha} and KM3NeT~\cite{Coniglione_2015}, are expected to detect one order of magnitude more astrophysical neutrinos. However, atmospheric neutrinos produced in EASs are an irreducible background to searches for astrophysical neutrinos. Ideally, one would eliminate any atmospheric background events, but this is experimentally not possible, instead, the remaining backgrounds must be estimated and subtracted. The acceptance for these events is typically determined using simulations and it is thus important to have precise experimental data to test these simulations and reduce their uncertainties. An accurate understanding of the physics of cosmic sources therefore requires an in-depth understanding of neutrino production in high-energy hadron interactions, in particular from heavy hadron decays. 

Atmospheric neutrinos are produced by semileptonic decays of hadrons in EAS, typically $\pi$ and $K$ decays (see also \cref{sec:light_hadrons}). This component is called the \emph{conventional} neutrino flux which falls steeply with increasing energy. However, when the energy is sufficiently high, atmospheric neutrinos also originate from the the semileptonic decay of heavy flavor hadrons, such as $D$ mesons, $B$ mesons, or $\Lambda_c$ baryons, for example. This component is called the \emph{prompt} neutrino flux. Due to their very short decay lengths, these hadrons decay immediately to neutrinos after they are produced and thus the flux of prompt neutrinos falls off less quickly with energy than the conventional neutrino flux. At high energies ($E_\nu\sim 10^5-10^6\,\rm{GeV}$), prompt atmospheric neutrinos are the main background to astrophysical neutrinos searches. For a neutrino energy of $10^6\,\rm{GeV}$, the corresponding center-of-mass energy is about $8\,\rm{TeV}$, a regime that can be probed at the HL-LHC.

\begin{figure}[tb]
\vspace{-1em}
\mbox{
\includegraphics[width=.49\linewidth]{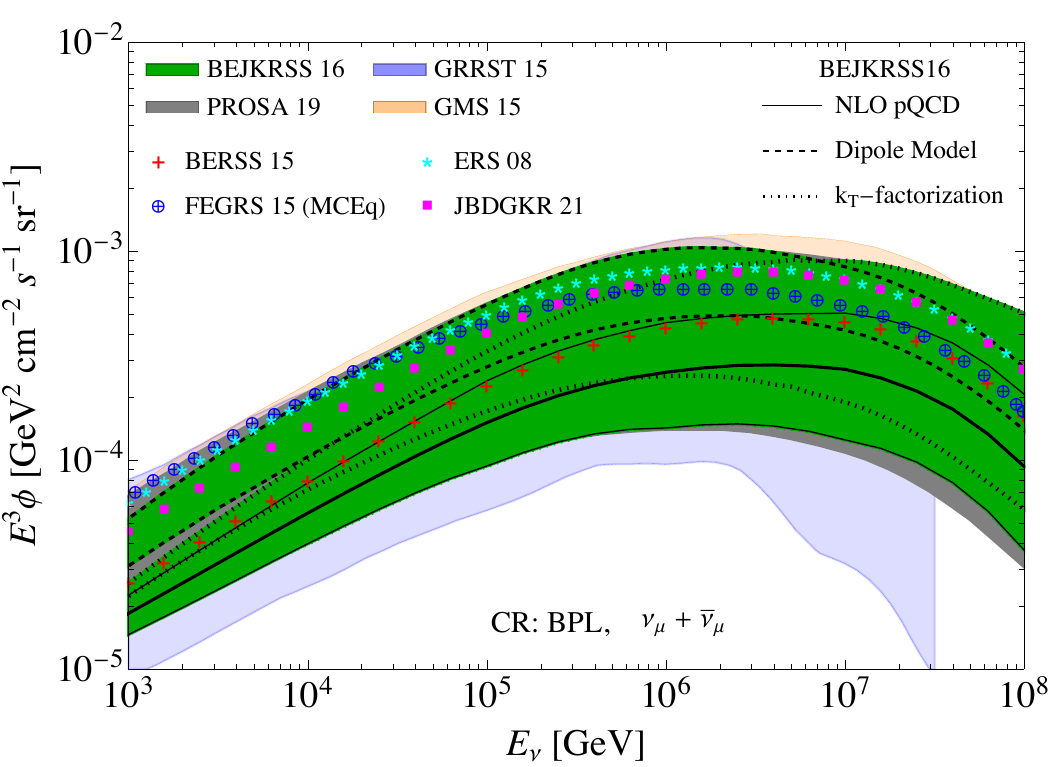}\,
\includegraphics[width=.49\linewidth]{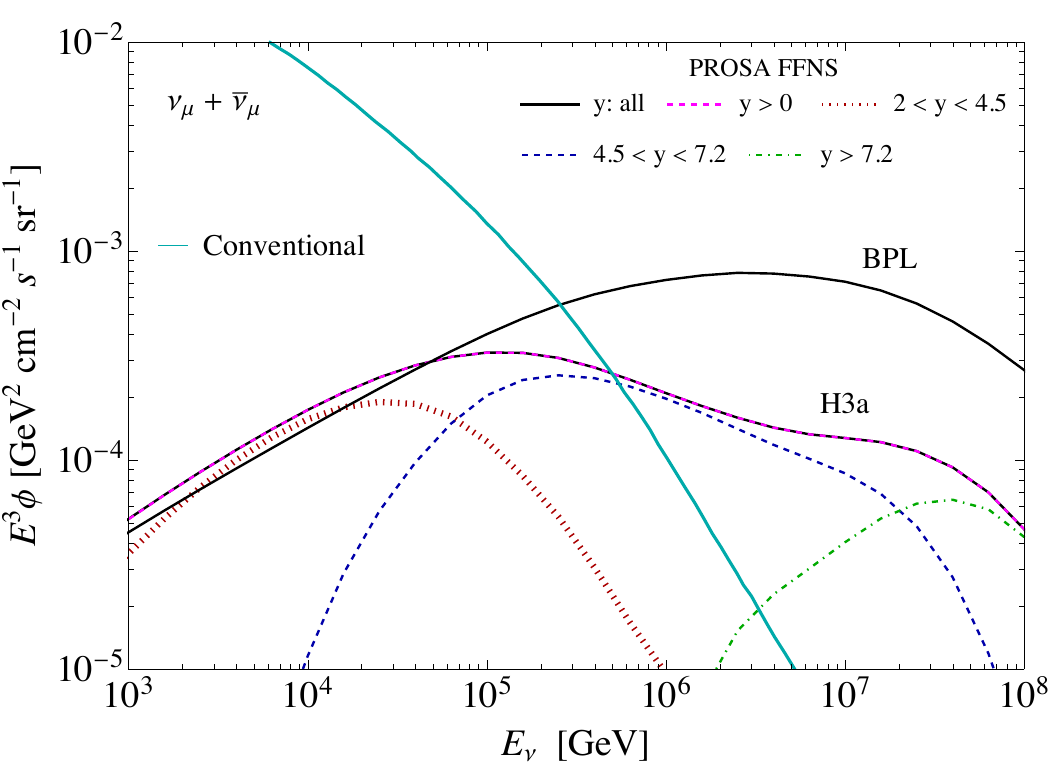}
}
\caption{\label{fig:PromptAtmNu}\textit{Left:} Comparison of the prompt atmospheric muon neutrino flux from various recent flux calculations, 
following Ref.~\cite{Zenaiev:2019ktw}. The incident cosmic ray flux is approximated with a broken power law in all predictions. If available, the accociated uncertainties are shown as error band. \textit{Right:} Prompt atmospheric muon neutrino flux produced in different collider rapidity ranges~\cite{Jeong:2021vqp}. Also shown is the conventional atmospheric neutrino flux from Ref.~\cite{Honda:2006qj}. The predictions are shown assuming a simple power law (BPL) for the initial cosmic ray flux and the H3a flux model from Ref.~\cite{Gaisser:2011klf}. Figure taken from Ref.~\cite{Feng:2022inv}.
}  
\end{figure}

While \cref{fig:nu-rate,fig:dos} show predictions of the charm contribution to the neutrino flux at the FPF obtained from different hadronic interaction models, in \cref{fig:PromptAtmNu} several modern predictions for the prompt atmospheric muon neutrino flux based on pQCD calculations are shown, following Ref.~\cite{Zenaiev:2019ktw}. These calculations assume a simple broken power law (BPL) cosmic ray spectrum, for a transparent comparison. As shown in the left panel of \cref{fig:PromptAtmNu}, the available theoretical predictions of the prompt atmospheric neutrino flux have very large uncertainties. These uncertainties originate from various input parameters in the evaluation of the prompt fluxes, e.g., the cross-sections for heavy flavor production, parton distribution functions, and fragmentation functions. As shown in the right panel, there is an additional, very large uncertainty from the cosmic ray flux assumption. \cref{fig:PromptAtmNu} (right) also shows the individual contributions from charm hadrons produced at different collider rapidity ranges. As shown in the figure, the prompt atmospheric neutrino flux, at energies where the prompt component is dominant, originates from charm hadrons produced at rapidities of $y > 4.5$, accessible by the proposed FPF experiments at the HL-HLC (see also Ref.~\cite{Jeong:2023gla}).

The FPF will be able to obtain data with unprecedented statistics
in the far-forward region, at energies relevant for astrophysical neutrino searches, thereby providing important information to reduce the uncertainties and strongly constrain models of charm hadron production. Consequently, the FPF will play a crucial role in improving the predictions of the prompt atmospheric neutrino flux. This, in turn, will have significant impact on the searches for astrophysical neutrinos by reducing the associated uncertainties.



\section{Conclusions}

The experiments at the proposed Forward Physics Facility at the HL-LHC will provide unique neutrino measurements with large statistics in the far-forward region. These measurements will probe hadron production in a phase space that is not accessible by any existing collider experiment but which is crucial for the modeling of high-energy hadronic interactions in the atmosphere. Hence, the measurements at the FPF have strong connections with modern astroparticle physics. They will reduce the uncertainties of air shower observations and in searches of astrophysical neutrinos, and thereby contribute to understanding the origin and nature of the highest energetic cosmic rays in the context of future multi-messenger observations~\cite{Coleman:2022abf}. 

\newpage 


\begin{spacing}{.7}
\bibliographystyle{ICRC}
\bibliography{references}
\end{spacing}

%
%
%

\end{document}